\documentclass[10pt]{article}
\usepackage[numbers]{natbib}
\usepackage{amsmath}
\usepackage{amsthm}
\usepackage{graphicx} 
\usepackage{natbib}
\usepackage{pifont}
\usepackage{array} 
\usepackage{comment} 
\usepackage{multirow}
\usepackage[table]{xcolor}
\usepackage{colortbl}
\usepackage{subcaption}
\usepackage{longtable}
\usepackage{enumitem}
\usepackage{url}
\usepackage{changepage}  
\usepackage{xurl} 
\usepackage[T1]{fontenc}
\usepackage{babel}
\usepackage{authblk}

\title{Can LLMs substitute SQL? Comparing Resource Utilization of Querying LLMs versus Traditional Relational Databases}

\author[1]{Xiang Zhang}
\author[2]{Khatoon Khedri}
\author[1]{Reza Rawassizadeh}
\affil[1]{Metropolitan College, Department of Computer Science, Boston University}
\affil[2]{Independent Scientist}
{
    \makeatletter
    \renewcommand\AB@affilsepx{: \protect\Affilfont}
    \makeatother

    \affil[ ]{Email ids}

    \makeatletter
    \renewcommand\AB@affilsepx{, \protect\Affilfont}
    \makeatother

    \affil[1]{xz0224@bu.edu}
    \affil[2]{khatoon.khedri1985@gmail.com}
    \affil[1,3]{rezar@bu.edu}
}

\begin{document}
\maketitle
\date{}

\begin{abstract}
Large Language Models (LLMs) can automate or substitute different types of tasks in the software engineering process. This study evaluates the resource utilization and accuracy of LLM in interpreting and executing natural language queries against traditional SQL within relational database management systems. We empirically examine the resource utilization and accuracy of nine LLMs varying from 7 to 34 Billion parameters, including Llama2 7B, Llama2 13B, Mistral, Mixtral, Optimus-7B, SUS-chat-34B, platypus-yi-34b,  NeuralHermes-2.5-Mistral-7B and Starling-LM-7B-alpha, using a small transaction dataset. Our findings indicate that using LLMs for database queries incurs significant energy overhead (even small and quantized models), making it an environmentally unfriendly approach. Therefore, we advise against replacing relational databases with LLMs due to their substantial resource utilization.

\end{abstract}

\section{Introduction}
The advent of Large Language Models (LLMs) has revolutionized several scientific and engineering disciplines, including software development tasks. Many software development related tasks could be done or automatized by LLMs. The satisfactory performance of LLM in search and query led to the introduction of specific LLM databases such as Vector database \cite{Zhang2023} auxiliary knowledge information retrieval methods, a.k.a., and Retrieval Augmented Generation \cite{Shao2023}.

Relational databases are one of the oldest and most common components of software applications. These databases manage structured data using interconnected tables in tabular form. Structured Query Language (SQL) is the query language used to interact with relational databases. 

There are two widely known and significant limitations of using LLMs: (i) factual mistakes and hallucinations caused by neural networks \cite{Tian2023}, and (ii) token size limitations \cite{Hoffmann2022}, which do not allow them to load a large dataset into their prompt, and thus have a limited data size. There are ongoing efforts to prove that the factuality and coverage of LLMs are quickly improving with new training architectures and the increasing amount of text used as input \cite{Elazar2021, Tam2022}. Besides, there are continuous efforts to increase or remove the token size limitation, such as using Structured state space models (S4), e.g., Mamba \cite{gu2023} instead of Transformers. 

Our work does not quantify or tackle any of these two known challenges. It focuses on benchmarking resource utilization using LLM instead of traditional SQL. In this research, we intend to investigate whether LLMs could replace traditional database management systems to search and query tabular data. Assuming even though the capability to generate SQL queries exists in LLMs, we should measure resource consumption and how accurately it identifies the correct answers from tabular datasets. 
 
An essential consideration in our exploration is the environmental impact of LLMs. There are ongoing discussions \footnote{\url{https://www.theatlantic.com/technology/archive/2024/03/ai-water-climate-microsoft/677602}}\footnote{\url{https://www.oregonlive.com/silicon-forest/2022/12/googles-water-use-is-soaring-in-the-dalles-records-show-with-two-more-data-centers-to-come.html}}\footnote{\url{https://www.bloomberg.com/news/articles/2023-07-26/thames-water-considers-restricting-flow-to-london-data-centers}}\footnote{\url{https://www.washingtonpost.com/business/2024/03/07/ai-data-centers-power}} on the huge electricity and water cooling supply, underscoring sustainability-related challenges brought about by the new existence and overall being of the LLMs. Our results testify that even using a small-size trained LLM still consumes a high amount of energy in comparison to a native SQL engine running on a relational database. Besides, we have observed the inferior accuracy of LLMs in comparison to SQL engines. However, larger models might resolve the accuracy problem in the near future, but the energy issue remains open.  

\section{Literature review}

There are recent reports on the water and electricity consumption of Generative Artificial Intelligence (AI) models, especially LLMs. However, their approach is mostly holistic and does not provide a comparative analysis of doing a particular task with LLM and without LLM \cite{Dodge2022,deVries2023,Luccioni2023, Li2023}. On the other hand, interest in adopting LLMs for general tasks like database querying has grown in the natural language processing community; there are several promising works in this direction, which we have categorized into two main groups. One group of work passes the query in natural language and data into an LLM and, as a result, gets the SQL query back. These works are known as \textit{Text-to-SQL} \citep{Xu2019,Tang2021,Wang2019,Baig2022,Ferreira2020}. The latter group \citep{Rawassizadeh2023, Deutch2017} provides the data and the query in natural language as input into an LLM. Then, they get the result in natural language as well, we call them \textit{NLQuery-to-NLAnswer}. In this section, we briefly describe each group of work.

\subsection{Text-to-SQL approaches}
Text-to-SQL approaches focus on transforming natural language queries into structured SQL commands, enabling users to interact with databases without needing SQL knowledge. The introduction of Google's SQL-PaLM model \citep{Sun2023} marks a pivotal development in natural language to SQL translation. SQL-PaLM model efficiently refines LLMs to understand the natural language query and convert it into SQL commands.

Baig et al. \citep{Baig2022} reviewed existing frameworks for processing natural language to SQL queries. The use of the attention mechanisms in neural networks for natural language interfaces to databases (NLIDB) was evaluated by Ferreira et al. \citep{Ferreira2020}. Wang et al. \citep{Wang2019} proposed the RAT-SQL framework, based on the relation-aware self-attention mechanism, to address schema encoding, schema linking, and feature representation within a Text-to-SQL encoder. RAT-SQL modeled the database schema as a directed graph. NADAQ \citep{Xu2019} merged specialized encoder-decoder architecture with traditional database parsing techniques for querying databases using natural language. 

\subsection{NLQuery-to-NLAnswer approaches}
Recently, Rawassizadeh and Rong \citep{Rawassizadeh2023} proposed ODSearch, which retrieves data from wearable and mobile devices through natural language processing. It employs data compression and Bloom filters to enable real-time responses to natural language queries.

Deutch et al. \citep{Deutch2017} presented a system that extends the generation of natural language interfaces to databases by generation of the natural language answer. It operates based on the provenance of the query result tuples. The provenance information is converted into natural language by structuring the originating query such that the user is delivered an informative response. Dries et al. \citep{Dries2009} also suggested a data model and query language designed specifically for network analysis in their research on a Query Language for Analysis Networks.

These works foster an interactive and less scripted interaction of a database system with the users. With those considerations, both Text-to-SQL and NLQuery-to-NLAnswer approaches highlight the importance of studying the resource usage of these systems. To our knowledge, except for ODSearch \citep{Rawassizadeh2023}, which does not use an LLM, none of the other works investigate the resource utilization of queries.

\subsection{Energy consumption of LLM}
Recently, the environmental impact of artificial intelligence has garnered significant attention from the research community, especially on water and electricity usage.

Large Language Models such as GPT-3 require substantial computational power for training, leading to significant Execution Energy Consumption and associated carbon emissions \footnote{\url{https://projectmanagers.net/top-10-disadvantages-of-large-language-models-llm}}.  Dodge et al. \citep{Dodge2022} present a method for calculating the carbon footprint of AI operations in the cloud, focusing on the energy consumption and CO2 emissions of machine learning models. The research highlights the importance of geographic location in selecting cloud instances to minimize carbon intensity.  Luccioni et al. \citep{Luccioni2023} conducted a systematic comparison of the energy and carbon costs associated with deploying various machine learning models. It reveals that multi-purpose, generative AI models, such as those used in LLM, are significantly more resource-intensive than task-specific models, even when accounting for model size. Their study calls for more intentional consideration of energy and emissions costs in the deployment of AI tools. De Vries \citep{deVries2023} explores AI's electricity use, considering both pessimistic and optimistic scenarios for global data center electricity consumption, and emphasizes the need for cautious adoption of AI technologies and understanding their energy implications. In addition to studies focused on energy utilization, Li et al. \citep{Li2023} examine the often-overlooked water footprint of AI corporations, particularly the substantial freshwater consumption by LLM models like GPT-3 during training in data centers. They estimate that global AI demand could lead to significant water withdrawal by 2027, emphasizing the urgency of addressing AI's water use. 

The most related works to ours are proposed by Tang et al. \citep{Tang2021}. They use machine learning to estimate SQL queries' CPU and memory demands, broadening evaluation beyond accuracy to include resource consumption, which is crucial for assessing LLMs' efficiency in database queries.

\section{Methodology}
In this work, we evaluate nine open-source LLMs that operate as NLQuery-to-NLAnswer. In particular, we measure their accuracy and resource utilization compared to SQL queries. 
Our study assesses how effectively LLMs are generating not only SQL but also direct answers from natural language queries. As an SQL engine, we choose to use (SQLite) \footnote{\url{https://sqlite.org/index.html}}, which is a common SQL engine used in devices that have resource constraints, such as Android phones.   There are promising tools available to measure the resource utilization of LLMs \citep{Samsi2023, MLEnergy}. However, we have used our scripts to have enough flexibility to measure different resources\footnote{\url{https://github.com/XiangZhang-zx/LLM-StockQuery-Dataset/blob/main/LLM_Generation_Comparison.ipynb}}. 

\subsection{Test Dataset}
The dataset is a synthetic representation of stock transactions in a real-world scenario built by SQLite\footnote{\url{https://github.com/XiangZhang-zx/LLM-StockQuery-Dataset/blob/main/dataset.csv}}.  SQLite's efficiency and minimal resource requirements make it suitable for scenarios where computational resources are limited, such as on battery-powered devices \citep{Rawassizadeh2023}. The synthetic dataset we built comprises 100 records across five stock symbols, such as AAPL, GOOGL, AMZN, MSFT, and TSLA, with the transaction type being BUY or SELL. Date of transactions, type, stock symbol, amount, and cost data attributes used in our queries. The transaction date was extracted along with its time from a series that spanned over a range of dates for seven consecutive days. Due to the small size of the test dataset, we do not encounter the token size limitation issue of LLM.

Amounts and costs were randomized using \textit{random} library to create a more realistic and diverse dataset \footnote{\url{https://docs.python.org/3/library/random.html}}. Instead of structuring our dataset with a schema, we directly feed 100 records into our framework. This decision reflected the more dynamic, real-world conditions under which non-expert users might interact with databases.
Based on the foundational concepts presented in \textit{Fundamentals of Database Systems} \cite{elmasri2016fundamentals}, the following are ten SQL queries designed to assess querying capabilities. These queries utilize \texttt{COUNT}, \texttt{SUM}, \texttt{MAX}, and \texttt{AVG}, apply condition filtering using \texttt{WHERE}, and implement grouping with \texttt{GROUP BY}.

\begin{enumerate}[label=\Alph*.]
    \item Count transactions per stock symbol.
    \item Total quantity sold per symbol.
    \item Total revenue from sales.
    \item Maximum sale price per symbol.
    \item Average purchase price per symbol.
    \item Several unique stock symbols.
    \item Quantities bought and sold per symbol.
    \item Total investment in buy transactions.
    \item The transaction quantity is on a specific date (2023-9-23).
    \item The highest transaction price for a stock on a specific date (google, 2023-9-24).
\end{enumerate}

\subsection{Experimental LLMs}
As shown in Table~\ref{tab:Parameters}, in our evaluation, we specifically chose a selection of large language models (LLMs), including Llama2 (7B and 13B versions), Mistral, and Mixtral, Optimus-7B, SUS-chat-34B, platypus-yi-34b, NeuralHermes-2.5-Mistral-7B, and Starling-LM-7B-alpha. These models were chosen based on ranking at the top of the Huggingface open LLM leaderboard (back in late 2023), and also our infrastructure can execute them. The traditional transformer stack was already designed to adapt them in terms of performance and efficiency. For Llama2 (7B and 13B), SUS-chat-34B, and platypus-yi-34b, we adhere to the traditional transformer stack. For Mistral, Mixtral, Optimus-7B, NeuralHermes-2.5-Mistral-7B, and Starling-LM-7B-alpha, we adhere to the pipeline produced by Hugging Face, tuned to a quantized 4-bit configuration.
\begin{table}[!ht]
    \centering
    \caption{Comparison of Large Language Models by Parameters and Configuration}
    \label{tab:Parameters}
    \hspace*{-4mm}
    \begin{tabular}{|l|l|l|}
    \hline
        Model  & Parameters & Configuration \\ \hline
        Llama2 7B & 7 Billion & Traditional Transformer \\ \hline
        Llama2 13B & 13 Billion & Traditional Transformer \\ \hline
        Mistral & 7 Billion & Hugging Face Pipeline, 4-bit Quantized \\ \hline
        Mixtral & 7 Billion & Hugging Face Pipeline, 4-bit Quantized \\ \hline
        Optimus-7B & 7 Billion & Hugging Face Pipeline, 4-bit Quantized \\ \hline
        SUS-chat-34B & 34 Billion & Traditional Transformer \\ \hline
        platypus-yi-34b & 34 Billion & Traditional Transformer \\ \hline
        NeuralHermes-2.5-Mistral-7B & 7 Billion & Hugging Face Pipeline, 4-bit Quantized \\ \hline
        Starling-LM-7B-alpha & 7 Billion & Hugging Face Pipeline, 4-bit Quantized \\ \hline
    \end{tabular}
\end{table}

\subsection{Experiment Setup}
Our hardware infrastructure includes two  NVidia RTX 4090 GPU 24GB, with 256 GB RAM and 3.30 GHz Intel Core i9 CPU. The operating system is  Ubuntu 20.04 LTS, and we used CUDA Version 12.0 for GPU computations.

To evaluate the performance of the LLM, we implemented a custom Python function that automates the process of measuring the time, CPU, and memory usage of the model. The function records these metrics before and after the model generates responses on natural language input using the \textit{tracemalloc} and \textit{time} libraries\footnote{\url{https://docs.python.org/3/library/tracemalloc.html}}\footnote{\url{ https://docs.python.org/3/library/time.html}}. Then, our function calculates the differences between the start and end values of the metrics and reports the execution time, CPU utilization, and memory consumption. To quantify energy consumption per process, the \textit{Turbostat} utility was employed to monitor the \textit{pkgwatt} (package power)\footnote{\url{https://www.linux.org/docs/man8/turbostat.html}}. This package, combined with the execution time, was used to calculate the model’s energy consumption in Joule (J).

In our experiments we use two pipelines, the Transformers Pipeline allows explicitly setting text generation performance and relevance with \textit{torch} library, combined with options control on temperature, max\_new\_tokens, as well as repetition\_penalty values\footnote{\url{https://pytorch.org/docs/stable/index.html}}. The Hugging Face Pipeline contains quantized models to reduce resource consumption using different options impacting output response sharpness and speed, such as max\_new\_tokens, top\_k, and eos\_token\_id values.

\section{Experimental Evaluation}
We examine the resource usage of SQL engine compared to  LLMS to query tabular data, the proficiency of LLMs in generating SQL-equivalent queries from natural language, and their effectiveness in obtaining semantically accurate responses from structured datasets.

To establish a baseline for the evaluation of LLMs, we measure both the execution time and memory consumption for queries (A-J) associated with direct SQL query execution. Based on our measurement of the direct SQL query execution on the SQL engine, the average execution Time is \textit{0.41 ms}, and the average memory usage is \textit{1641 B.}  As we have described earlier, the SQL engine we used is SQLite.

The average execution time and memory utilization for direct query results and query generation of LLM models are presented in Tables ~\ref{tab:performance_matrices_direct} and ~\ref{tab:performance_matrices_SQL}. Moreover, we display the accuracy of direct query results by LLM models in Table~\ref{tab:overall-accuracy} and the overall accuracy of them in Table~\ref{tab:overall-accuracy-SQL}.

In the results shown in Tables \ref{tab:overall-accuracy} and Tables \ref{tab:overall-accuracy-SQL}, symbols used are \ding{51} for correct generation and \ding{55} for incorrect or incomplete generation.
\subsection{Natural Language Query Performance Analysis}

We present the results of our comparison by focusing on different aspects of the models, including execution time and accuracy. As shown in Table~\ref{tab:performance_matrices_direct}, the average execution time varied significantly across the models, from as quick as 23 seconds for Mistral to 260 seconds for SUS-chat-34B. It indicates that the size and architecture of the models have a significant impact on the execution of the tasks. SUS-chat-34B also showed the highest memory usage in the transformer pipeline, highlighting the scalability concerns of using large and complex models for natural language processing tasks. Notably, in the Hugging Face pipeline, models like Optimus-7B demonstrated efficiency with minimal memory increase, proving that using quantization techniques can reduce the resource consumption of the models. Our results suggest that larger LLMs can achieve higher accuracy for natural language processing tasks but also pose challenges in terms of execution time and resource utilization.

According to Table~\ref{tab:performance_matrices_direct}, Llama2 7B was the most resource-efficient model across the tasks, with reasonable execution times and resource usage. SUS-chat-34B, on the other hand, had high resource consumption, raising questions about its practicality in larger datasets. Optimus-7B, which employs quantization techniques to reduce model size and complexity, comes closest to achieving the execution time and resource efficiency of SQLite. 

In Table~\ref{tab:performance_matrices_direct}, platypus-yi-34b accurately interpreted straightforward queries, such as identifying the total number of unique stock symbols. However, models often predict or complete questions rather than providing the requested information, highlighting a propensity for these models to engage in dialogue rather than execute database queries accurately. Regarding inconsistencies, Llama2 7B and Llama2 13B sometimes generated irrelevant responses, indicating a need for improved training focused on database querying capabilities.

\begin{table}[!ht]
    \centering
    \caption{Average execution time and memory utilization of direct query results of LLM models}
    \begin{tabular}{lll}
    \hline
        Model & Execution Time (s) & Memory Usage (kB) \\ \hline\hline
        Llama2 7B & 60 & 64 \\ 
        Llama2 13B & 106 & 70 \\ 
        SUS-Chat-34B & \cellcolor{green!25}260 & 63 \\ 
        platypus-yi-34b & 235 & 70 \\ 
        Mistral & \cellcolor{red!25}23 & 301 \\ 
        NeuralHermes-2.5-Mistral-7B & 78 & 464 \\ 
        Optimus-7B & 33 & 247 \\ 
        Starling-LM-7B-alpha & 41 & 263 \\ 
        Mixtral & 116 & 571 \\ \hline 
    \end{tabular}
    \label{tab:performance_matrices_direct}
\end{table}

\bigskip

\begin{table}[!ht]
    \centering
    \caption{Accuracy of direct Query Results of LLM Models} 
    \footnotesize 
    \label{tab:overall-accuracy}
    \hspace*{-7mm} 
    \begin{tabular}{llllllllll}
    \hline

        Model & 
\begin{tabular}[c]{@{}l@{}}Llama\\2\\ 7B\end{tabular} & 
\begin{tabular}[c]{@{}l@{}}Llama\\2\\ 13B\end{tabular} & 
\begin{tabular}[c]{@{}l@{}}SUS-\\Chat\\ 34B\end{tabular} & 
\begin{tabular}[c]{@{}l@{}}platy\\pus-yi\\ -34b\end{tabular} & 
Mistral & 
\begin{tabular}[c]{@{}l@{}}Neural\\Hermes\\ -2.5-Mistral\\ -7B\end{tabular} & 
\begin{tabular}[c]{@{}l@{}}Optimus\\ -7B\end{tabular} & 
\begin{tabular}[c]{@{}l@{}}Starling\\ -LM\\ -7B-alpha\end{tabular} & 
Mixtral \\ \hline\hline
        A & \ding{55} & \ding{55} & \ding{55} & \ding{55} & \ding{55} & \ding{55} & \ding{55} & \ding{55} & \ding{55} \\ 
        B & \ding{55} & \ding{55} & \ding{55} & \ding{55} & \ding{55} & \ding{55} & \ding{55} & \ding{55} & \ding{55} \\ 
        C & \ding{55} & \ding{55} & \ding{55} & \ding{55} & \ding{55} & \ding{55} & \ding{55} & \ding{55} & \ding{55} \\ 
        D & \ding{55} & \ding{55} & \ding{55} & \ding{55} & \ding{55} & \ding{55} & \ding{55} & \ding{55} & \ding{55} \\ 
        E & \ding{55} & \ding{55} & \ding{55} & \ding{51} & \ding{55} & \ding{55} & \ding{55} & \ding{55} & \ding{55} \\ 
        F & \ding{55} & \ding{55} & \ding{55} & \ding{55} & \ding{55} & \ding{55} & \ding{55} & \ding{55} & \ding{55} \\ 
        G & \ding{55} & \ding{55} & \ding{55} & \ding{55} & \ding{55} & \ding{55} & \ding{55} & \ding{55} & \ding{55} \\ 
        H & \ding{55} & \ding{55} & \ding{55} & \ding{55} & \ding{55} & \ding{55} & \ding{55} & \ding{55} & \ding{55} \\ 
        I & \ding{55} & \ding{55} & \ding{55} & \ding{55} & \ding{55} & \ding{55} & \ding{55} & \ding{55} & \ding{55} \\ 
        J & \ding{55} & \ding{55} & \ding{55} & \ding{55} & \ding{55} & \ding{55} & \ding{55} & \ding{55} & \ding{55} \\\hline
        Accuracy & 0\% & 0\% & 0\% & 10\% & 0\% & 0\% & 0\% & 0\% & 0\% \\ \hline
    \end{tabular}
    \smallskip
\end{table}

\subsection{SQL Query Generation Results}

We evaluated listed LLMs for generating SQL queries from natural language inputs, and Table~\ref{tab:performance_matrices_SQL}  displays the average execution time and memory utilization of SQL query generation using our experimental LLMs.

Llama2 7B, Llama2 13B, and Mistral 7B showed mixed results in translating natural language to SQL, ranging from partially accurate to essentially reiterating the initial query. Another important observation from the experiments was that most of the models, including Mistral 7B, SUS-Chat-34B, platypus-yi-34b, Optimus-7B, and Starling-LM-7B-alpha, failed to include the condition of transaction is equal to SELL or BUY in their SQL queries.  Table~\ref{tab:performance_matrices_SQL} shows that in the transformer pipeline,  while SUS-chat-34B and platypus-yi-34b demonstrated high success in correct script generation, but their high resource consumption is a challenge. Conversely, within the Hugging Face pipeline, Optimus-7B and Starling-LM-7B-alpha achieved accurate SQL generation with lower resources.

Table~\ref{tab:overall-accuracy-SQL} shows meaningful variability in model performance, with some models excelling in accuracy while others struggled with resource utilization and generating precise SQL queries. 

\subsection{Energy Utilization}
\begin{figure}[htbp]
  \centering
  \includegraphics[width=\linewidth]{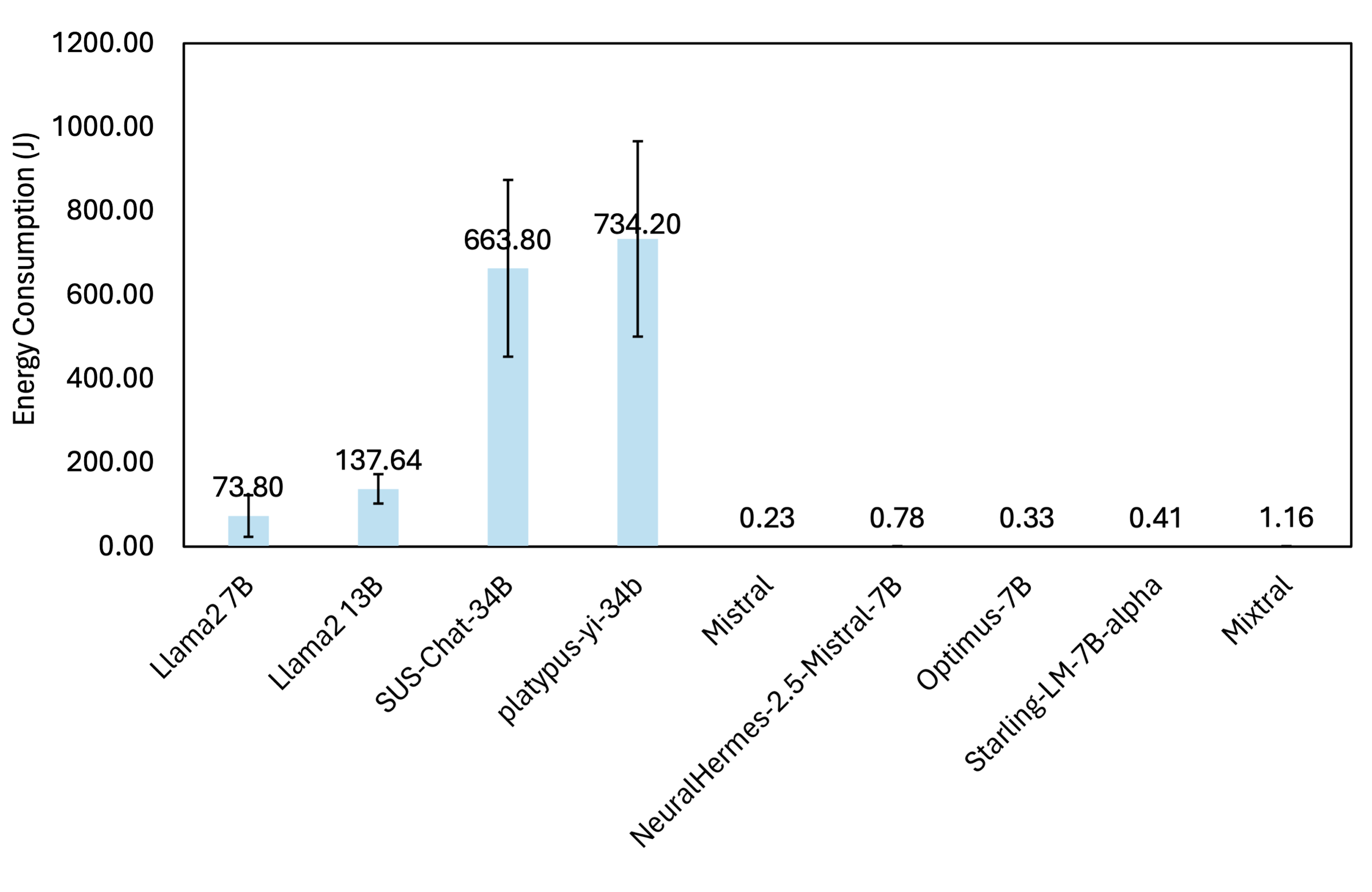}
  \caption{The average energy consumption (J) for direct query results of LLM models}
  \label{fig:direct_generation_energy}
\end{figure}

\begin{figure}[htbp]
  \centering
  \includegraphics[width=\linewidth]{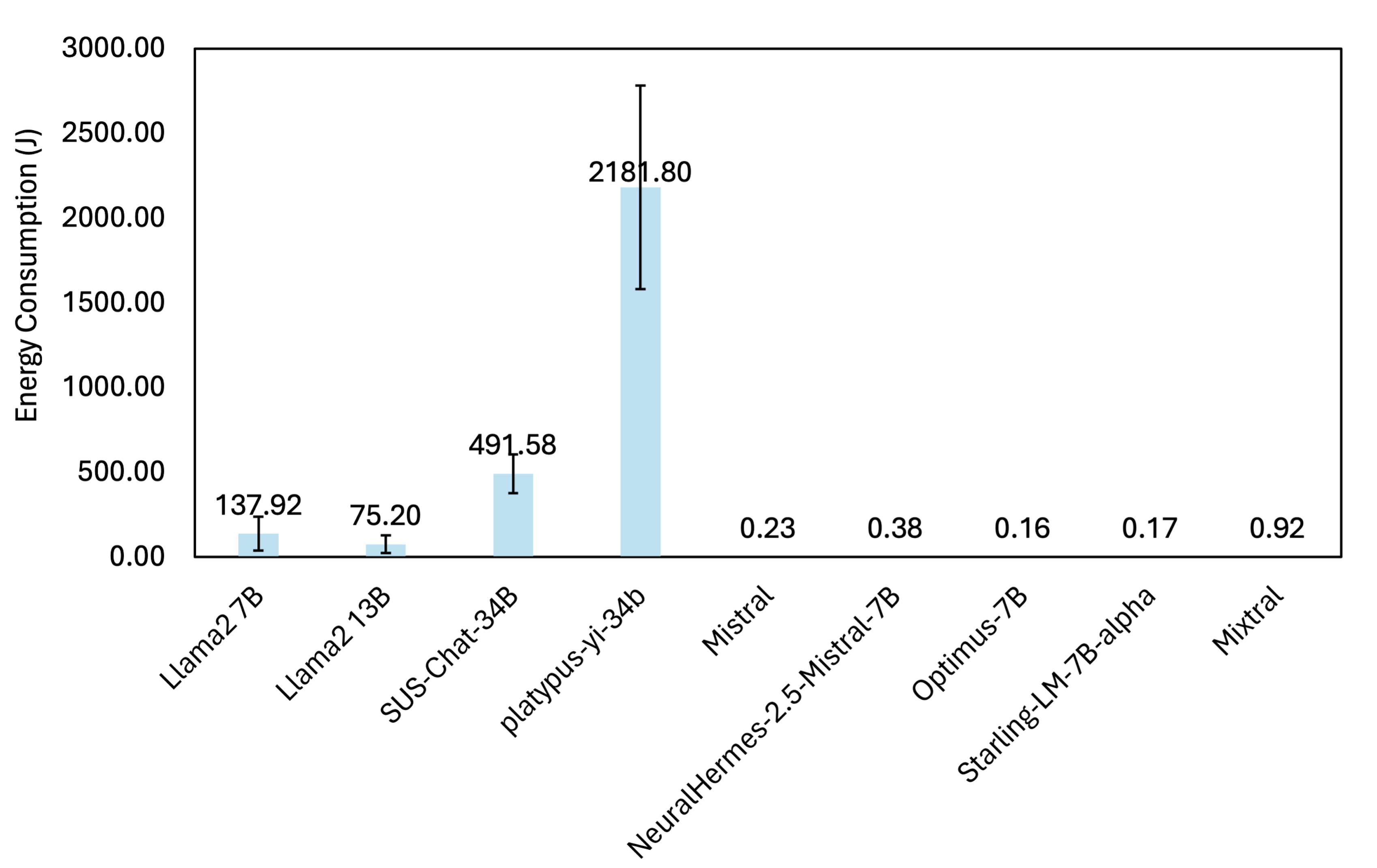}
  \caption{The average energy consumption (J) for SQL query generation of LLM models}
  \label{fig:sql_generation_energy_llm}
\end{figure}

Figures \ref{fig:direct_generation_energy} and \ref{fig:sql_generation_energy_llm} present the average energy utilization for direct SQL query execution along LLM models. We can observe that SQL engine consumes the least energy, quantified at 8.22$\times10^{-6}$J. In the assessment of LLM models for both direct query execution and SQL query generation, Platypus-yi-34b was identified as the most energy-intensive, recording energy utilization of 2181.8J and 734.2J, respectively. In contrast, Optimus-7B exhibited the lowest energy consumption for direct query execution at 0.163J, while Mistral registered the lowest for SQL query generation, consuming 0.234J. Therefore, we can conclude that the larger the model, the more utilized energy is used to run a query.


\begin{table}[!ht]
    \centering
    \caption{Average execution time and memory utilization of SQL query generation using LLM models}
    \begin{tabular}{lll}
    \hline
        Model & Execution Time (s) & Memory Usage (kB) \\ \hline \hline
        Llama2 7B & 106 & 70 \\ 
        Llama2 13B & 61 & 55 \\ 
        Mistral & 23 & 232 \\ 
        SUS-Chat-34B & 200 & 57 \\ 
        platypus-yi-34b & 597 & 93 \\ 
        NeuralHermes-2.5-Mistral-7B & 38 & 266 \\ 
        Optimus-7B & 16 & 206 \\ 
        Starling-LM-7B-alpha & 17 & 204 \\ 
        Mixtral & 92 & 488 \\ \hline
    \end{tabular}
    \label{tab:performance_matrices_SQL}
\end{table}

\begin{table}[!ht]
    \centering
    \caption{Detailed Accuracy of Query Generation} 
    \label{tab:overall-accuracy-SQL}
  \footnotesize 
      \hspace*{-6mm} 
    \begin{tabular}{llllllllll}
    
     \hline
        Model &
\begin{tabular}[c]{@{}l@{}}Llama\\2\\ 7B\end{tabular} & 
\begin{tabular}[c]{@{}l@{}}Llama\\2\\ 13B\end{tabular} & 
\begin{tabular}[c]{@{}l@{}}SUS-\\Chat\\ 34B\end{tabular} & 
\begin{tabular}[c]{@{}l@{}}platy\\pus-yi\\ -34b\end{tabular} & 
Mistral & 
\begin{tabular}[c]{@{}l@{}}Neural\\Hermes\\ -2.5-Mistral\\ -7B\end{tabular} & 
\begin{tabular}[c]{@{}l@{}}Optimus\\ -7B\end{tabular} & 
\begin{tabular}[c]{@{}l@{}}Starling\\ -LM\\ -7B-alpha\end{tabular} & 
Mixtral \\ \hline\hline
        A & \ding{55} & \ding{51} & \ding{51} & \ding{51} & \ding{55} & \ding{55} & \ding{51} & \ding{51} & \ding{55} \\ 
        B & \ding{55} & \ding{55} & \ding{51} & \ding{51} & \ding{51} & \ding{55} & \ding{51} & \ding{51} & \ding{51} \\ 
        C & \ding{51} & \ding{55} & \ding{51} & \ding{55} & \ding{55} & \ding{55} & \ding{55} & \ding{51} & \ding{51} \\ 
        D & \ding{55} & \ding{55} & \ding{55} & \ding{55} & \ding{55} & \ding{55} & \ding{55} & \ding{55} & \ding{51} \\ 
        E & \ding{55} & \ding{55} & \ding{51} & \ding{55} & \ding{55} & \ding{55} & \ding{55} & \ding{55} & \ding{55} \\ 
        F & \ding{55} & \ding{55} & \ding{51} & \ding{51} & \ding{51} & \ding{51} & \ding{51} & \ding{51} & \ding{55} \\ 
        G & \ding{55} & \ding{55} & \ding{55} & \ding{51} & \ding{55} & \ding{51} & \ding{51} & \ding{51} & \ding{51} \\ 
        H & \ding{55} & \ding{55} & \ding{55} & \ding{55} & \ding{55} & \ding{55} & \ding{55} & \ding{55} & \ding{51} \\ 
        I & \ding{55} & \ding{55} & \ding{51} & \ding{55} & \ding{51} & \ding{55} & \ding{51} & \ding{55} & \ding{55} \\ 
        J & \ding{55} & \ding{51} & \ding{51} & \ding{51} & \ding{51} & \ding{51} & \ding{55} & \ding{51} & \ding{51} \\   \hline
        Accuracy & 10\% & 20\% & 70\% & 50\% & 40\% & 30\% & 50\% & 60\% & 60\% \\ \hline

    \end{tabular}
    \smallskip
\end{table}
\stepcounter{table}

\section{Discussion}
Direct query results from LLM models show disappointingly low accuracy. These findings highlight a significant challenge: LLMs struggle to query databases effectively without additional engineering. Specifically in generating SQL queries, Models often misinterpreted complex requests, incorrectly applying SQL clauses.

Our findings also point out that energy efficiency varies among LLM models used for SQL query generation, with larger models consuming more energy. Quantized models, such as Optimus-7B, performed well in the execution time and resource use, but limitations in scalability and token size question their efficacy on larger datasets. Nonetheless, LLMs could enhance DBMS querying alongside traditional methods, improving accessibility for non-experts. Further research should aim at hybrid methodologies that combine LLM capabilities with traditional SQL parsing technologies.

\section{Conclusion and Future Work}
LLMs offer a radically new perspective on database querying and the nature of computational systems. In this work, we measure the accuracy and resource utilization of nine small open-source LLMs in querying tabular data. Our results present the significant resource expense of employing LLMs, even small models that are highly compressed with quantization. Besides, the accuracy of using LLM (at least not the very large and commercialized ones) for querying tabular data is low. As the model gets larger, the accuracy improves, but we didn't experiment with larger models. Potential further research can investigate fine-tuning existing models with SQL schema, toward reducing the misinterpretations made by the LLM models in querying databases.

\bibliographystyle{plain}
\bibliography{references}

\end{document}